\documentclass[aps,pra,twocolumn, notitlepage,10pt,superscriptaddress,longbibliography]{revtex4-1}
\usepackage{graphicx}
\usepackage{times,bbm,amsmath,amssymb}
\usepackage{hyperref}
\usepackage{color}
\usepackage{xcolor}
\usepackage{graphicx}
\usepackage{dcolumn}%
\usepackage{bm}
\usepackage{soul}

\usepackage{nicefrac, xfrac}
\usepackage{amssymb}
\usepackage{pifont}

\usepackage{physics}
\begin{document}
\title{Twin-beam advantage in quantum LiDAR under correlated noise}

\author{Valeria Cimini}
\email{valeria.cimini@uniroma1.it}
\affiliation{Dipartimento di Fisica, Sapienza Universit\`{a} di Roma, Piazzale Aldo Moro 5, I-00185 Roma, Italy}

\begin{abstract}
Quantum light promises improved precision in optical remote sensing, but its practical advantage depends critically on whether nonclassical resources remain useful under realistic noise and experimentally accessible detection. This question becomes especially relevant for LiDAR systems, where a quantum advantage has been demonstrated for target detection and joint range–velocity estimation, but mostly under idealized conditions or simple noise models, such as optical loss and thermal background. A key open point is whether entanglement provides an operational advantage when the dominant disturbance is not independent noise, but structured interference across sensing modes.
Here, we address this question by studying the joint estimation of target range and velocity with bright two-mode Gaussian probes and homodyne detection, comparing coherent, separable squeezed, and twin-beam states at a fixed resource budget. Our results reveal a hierarchy of quantum resources set by the noise structure: separable squeezing provides a robust advantage over coherent illumination under loss and thermal background, whereas twin-beam probes become superior under correlated jamming when the receiver is adaptively optimized.
These results establish correlated noise as the operational regime in which entanglement provides a robustness advantage beyond local squeezing, opening a receiver-aware route to quantum-enhanced LiDAR in realistic and potentially adversarial environments.
\end{abstract}

\maketitle

Among quantum sensing platforms, light occupies a special position \cite{polino2020photonic, barbieri2022optical, pirandola2018advances}. Optical fields can be prepared in highly nonclassical states, manipulated, and measured with mature homodyne or heterodyne and photon-counting technologies. Squeezed light has already become a paradigmatic resource for quantum-enhanced interferometry \cite{caves1981quantum, minati2025multiparameter, lawrie2019quantum}. The use of quantum states of light has been vastly investigated, particularly in situations with stringent power constraints on the intensity of the light probe, such as the study of biological samples, where excessive optical power can damage delicate specimens \cite{taylor2016quantum, moreva2025quantum, cimini2019adaptive}. Beyond these local sensing settings, however, light can also be exploited in remote sensing applications, where electromagnetic radiation is used to acquire information about distant targets.  
When high spatial resolution and high-precision target ranging are required over shorter distances, optical wavelengths offer distinct advantages. Systems operating in this frequency range are known as Light Detection and Ranging (LiDAR) \cite{dong2017lidar, mcmanamon2012review, wang2021challenges, northend1966laser}.
Beyond target ranging and velocity estimation, LiDAR underpins atmospheric monitoring \cite{comeron2017current}, aerosol and gas sensing, topographic mapping \cite{amann2001laser}, autonomous navigation \cite{royo2019overview}, and environmental spectroscopy. In all these settings, the optical field has to be measured after propagation through a lossy, noisy, and often uncontrolled channel.

As LiDAR systems become increasingly deployed in complex and potentially adversarial scenarios, robustness to structured noise, spoofing, and intentional jamming is becoming as important as raw estimation precision ~\cite{cao2019adversarial,petit2015remote}. Whether a quantum probe retains its metrological advantage under such structured, possibly adversarial, classical interference is the central open question of modern quantum-enhanced LiDAR~\cite{karsa2024quantum, mrozowski2024demonstration, blakey2022quantum, chen2024optimal, zhuang2017entanglement, xu2021experimental, torrome2024advances, spedalieri2021optimal, zhuang2021quantum, sorelli2021detecting, manrique2026quantum}. 
Many existing protocols rely on idler storage or measurement strategies that are difficult to implement in practical remote-sensing architectures. \cite{lloyd2008enhanced, shapiro2009quantum, tan2008quantum, jo2021quantum, reichert2023quantum, giovannetti2001quantum}. At the same time, realistic sensing problems are naturally multiparametric: the returned field generally encodes several target properties simultaneously, making a multiparameter description essential for a faithful characterization of the sensing protocol \cite{huang2021quantum, zhuang2017entanglement, reichert2024heisenberg}. This has driven growing interest in quantum multiparameter estimation, both as a fundamental problem in metrology and as a framework for experimentally relevant sensing architectures \cite{cimini2024benchmarking, belliardo2024optimizing,cimini2024variational, valeri2023experimental, cimini2023deep, roccia2018multiparameter, minati2025multiparameter}. 
The central practical question is therefore not only whether quantum resources can improve precision in principle, but whether such an advantage survives under realistic operating conditions and remains accessible with experimentally implementable receivers \cite{zhang2015entanglement, li2025noise}.

\begin{figure*}[htb!]
	\includegraphics[width=0.99\textwidth]{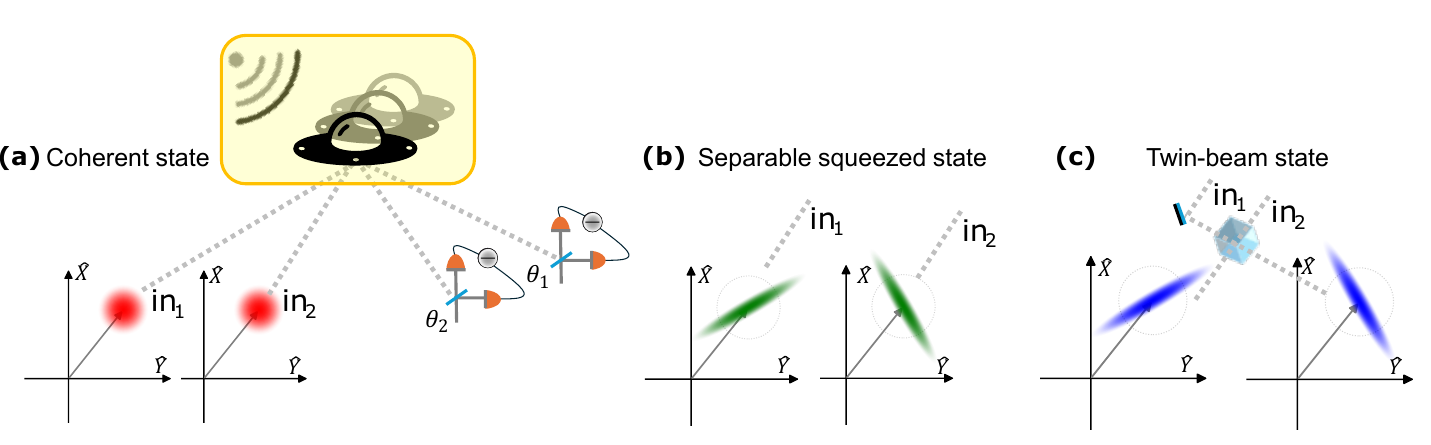}
	\caption{LiDAR scheme for the joint estimation of target range and velocity in a noisy environment. The protocol is investigated for three two-mode ($\text{in}_1$ and $\text{in}_2$) probe families: \textbf{(a)} coherent states, \textbf{(b)} separable squeezed states, and \textbf{(a)} twin-beam states. After target interaction and propagation through the noisy channel, the two reflected modes are analyzed by homodyne detection. The local-oscillator phases $\theta_1$ and $\theta_2$ can be optimized depending on the considered configuration.}
	\label{fig0}
\end{figure*}

In this work, we show that correlated noise defines a regime in which entanglement provides a structural advantage over local squeezing. We address this challenge by studying the joint estimation of a delay-like parameter $\tau$ associated with target range, and a frequency-like parameter $\omega$, associated with target velocity, using two-mode optical probes.
We compare three resource classes at a fixed resource budget: classical coherent states, separable squeezed states, and entangled twin-beam states. 
Crucially, our analysis is formulated directly at the receiver level, in a practically implementable setting that requires no idler storage, employs only homodyne detection (HD), and operates with bright two-mode nonclassical states compatible with experimentally available squeezing levels. We evaluate both Fisher information (FI) and explicit estimator performance from the quadratures actually measured by the receiver, including maximum-likelihood reconstruction from simulated homodyne data.

Our central result is that the quantum advantage has a structured hierarchy. In ideal conditions, separable squeezed probes outperform coherent probes, saturating the ultimate precision bound. 
The advantage of employing twin-beam states emerges in the presence of correlated jamming noise, where the disturbance itself has an intermodal structure. In that regime, the entangled probe can outperform the separable squeezed probe, if the receiver is allowed to adapt its measurement settings.
Our work moves from a state-level characterization of ultimate precision limits and develops an actually implementable model based on bright Gaussian probes, realistic loss, structured noise, and adaptive homodyne-based receivers.
These results, therefore, establish a practical, receiver-aware route to quantum-enhanced LiDAR, not only present at the level of ideal bounds, which remains accessible with experimentally accessible measurements and estimators under realistic conditions.

\emph{Two-mode Gaussian probes for joint range and velocity estimation --}
We consider a two-mode LiDAR probe following the idler-free configuration of \cite{reichert2024heisenberg}, in which both signal modes are directed toward the target and then measured at the receiver station.
We adopt a reduced description in terms of two sensing channels, encoding the two unknown parameters through independent phase-space rotations $\hat{R}_1(\Delta\Omega\tau)$ and $\hat{R}_2(\Delta T\omega)$, 
where $\tau$ is related to the delay in arrival time, $\omega$ to the Doppler frequency shift, and $\Delta\Omega$ and $\Delta T$ set the natural scales of the two encodings. This minimal model provides a transparent receiver-level setting for the analysis. As shown in the Appendix, the same resource hierarchy persists for more general two-mode encoding $\hat{R}_1(a\tau+b\omega)\otimes\hat{R}_2(c\tau+d\omega)$, confirming that the results obtained are independent of this specific choice.

Within this framework, we compare three classes of Gaussian probes, schematized in Fig.\ref{fig0}: a classical coherent probe, a quantum probe consisting of two separable displaced squeezed modes, and an entangled quantum probe based on twin-beam states. This allows us to distinguish the role of different quantum resources in the sensing task, separating local squeezing from the additional contribution of EPR correlations.

To ensure a fair comparison, all probe families are evaluated at fixed total signal energy. The same total photon number $N$ is distributed across the two sensing modes. In this way, any observed performance difference can be attributed to the structure of the probe correlations rather than to an unequal allocation of optical resources.

The classical probe consists of a two-mode coherent probe 
\begin{equation}
 |\psi\rangle_\text{cl} = |\alpha_1\rangle_1 \otimes |\alpha_2\rangle_2,   
\end{equation}
with coherent amplitudes $\alpha_1 = \sqrt{N/2} e^{i\phi_1}$ and $\alpha_2 = \sqrt{N/2} e^{i\phi_2}$, respectively.
To study the effect of squeezing on estimation precision, we compare this classical probe with two nonclassical Gaussian probes prepared under the same total photon-number constraint. In both quantum cases, the available signal energy is partitioned into a squeezed contribution $N_{\text{sq}} = f_{\text{sq}} N$ that sets the squeezing parameter to $r = \text{arcsinh}\sqrt{N_\text{sq}/2}$, and a coherent contribution $N_{\text{coh}} = (1-f_{\text{sq}}) N$, fixing $f_{\text{sq}}$, which allows us to maintain a realistic bright probe. To remain in an experimentally realistic regime, the analysis is carried out with $N=50$ and $f_{\text{sq}}=0.3$, ensuring that the squeezing-photon budget stays within currently attainable values ($r<1.8$).
The separable quantum benchmark consists of two separable squeezed bright signal modes:
\begin{equation}
    |\psi\rangle_\text{sep} = \hat{D}_1(\alpha_1) \hat{D}_2(\alpha_2)\hat{S}_1(r)\hat{S}_2(r)|0\rangle_1|0\rangle_2,
\end{equation}
where $\hat{D}$ denotes the displacement operator with amplitudes $\alpha_1 = \sqrt{N_\text{coh}/2} e^{i\phi_1}$ and $\alpha_2 = \sqrt{N_\text{coh}/2} e^{i\phi_2}$, and $\hat{S}=e^{\frac{1}{2}(r\hat{a}^2-r\hat{a}^{\dagger 2})}$ is the single-mode squeezing operator.
In this configuration, both sensing modes are squeezed along the same quadrature direction, while no entanglement is introduced between them. The probe thus contains genuine quantum resources, but only in the form of local single-mode squeezing. The parameter encoding is then applied through the local phase rotations described above.
The entangled probe is prepared as a bright two-mode squeezed state, obtained by applying a two-mode squeezing operation  $\hat{S2}=e^{(r\hat{a}^{\dagger}_1\hat{a}^{\dagger}_2-r\hat{a}_1\hat{a}_2)}$, that introduces genuine EPR correlations between the two modes:
\begin{equation}
|\psi\rangle_\text{twin} = \hat{D}_1(\alpha_1) \hat{D}_2(\alpha_2)\hat{S2}(r)|0\rangle_1|0\rangle_2.   
\end{equation}

\begin{figure*}[htb!]
	\includegraphics[width=0.99\textwidth]{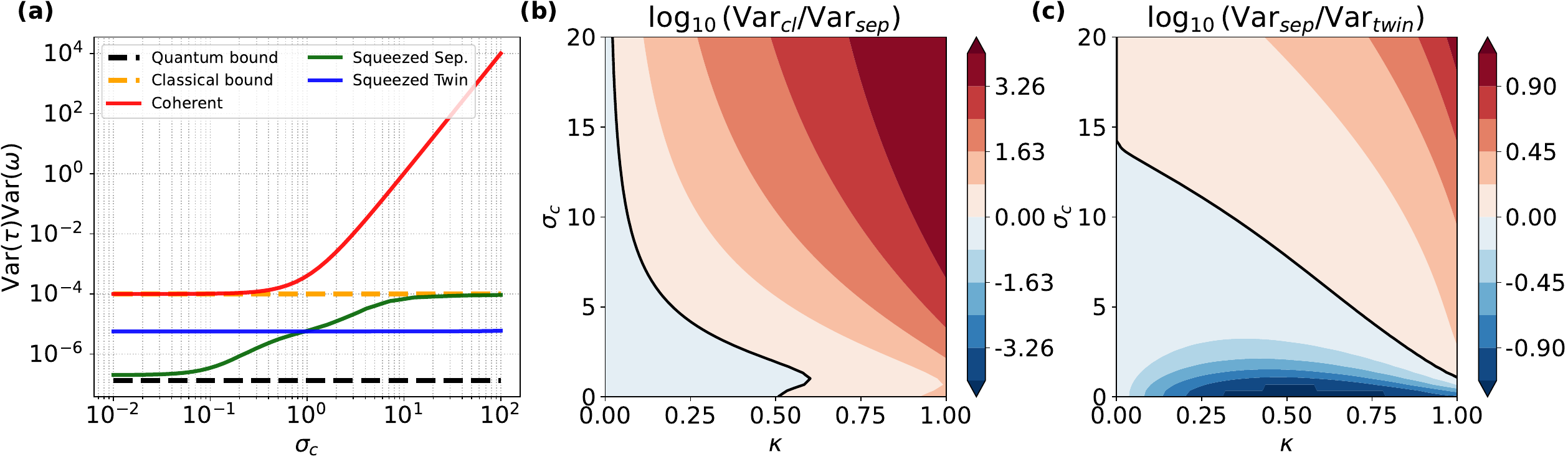}
	\caption{\textbf{(a)} Joint range–velocity estimation performance against correlated noise strength $\sigma_c$ at fixed probe energy $N=50$ for $n_\text{th} = 0$ and $\kappa=1$ obtained from homodyne quadratures for coherent states (red), separable squeezed states (green), and twin-beam states (blue). The standard quantum limit for classical light is the orange dashed line ($1/(4\Delta\Omega^2\Delta T^2N^2)$), while the ultimate Heisenberg limit is the black dashed line (($1/(4\Delta\Omega^2\Delta T^2N^4)$)). \textbf{(b)} Comparison of coherent states versus separable squeezed states for different levels of loss and correlated jamming noise strength at fixed thermal background $n_\text{th}=0.5$. The panels report the logarithm of the ratio between the variance products obtained by optimizing the LO phases for different probe classes, as a function of the channel transmissivity $\kappa$ and the strength $\sigma_c$ of the correlated classical noise. Positive values (red color map) indicate an advantage of the probe appearing in the denominator, while negative values (blue color map) indicate an advantage of the probe in the numerator; the black contour marks the crossover where the two strategies perform equally. \textbf{(c)} Comparison of separable squeezed states versus twin-beam states.}
	\label{fig1}
\end{figure*}

To compare different probe families within a common and experimentally relevant description, we cast the entire protocol in a unified Gaussian framework. As a result, the two-mode sensing problem admits a complete statistical description in terms of the quadrature vector $\mathbf{\hat{X}} =(\hat{x}_1,\hat{x}_2,\hat{p}_1,\hat{p}_2)^T$.
The returned optical state is therefore fully characterized by the mean vector $\mathbf{\overline{x}} = \langle\mathbf{\hat{X}}\rangle$ and the covariance matrix $\mathbf{\Sigma}_{ij} = \frac{1}{2} \langle\hat{X}_i\hat{X}_j+\hat{X}_j\hat{X}_i\rangle - \langle\hat{X}_i\rangle\langle\hat{X}_j\rangle$, both functions of the parameters of interest.
The estimation precision is then linked to the Fisher information matrix (FIM) \cite{bressanini2024multi, nichols2018multiparameter} given by:

\begin{equation}
\begin{split}
    \mathbf{F}_{\alpha\beta} &= (\frac{\partial}{\partial\theta_\alpha}\overline{\mathbf{x}}^T)\mathbf{\Sigma}^{-1}(\frac{\partial}{\partial\theta_\beta}\overline{\mathbf{x}}) +\\ 
    &+\frac{1}{2} \text{Tr}\big[ \mathbf{\Sigma}^{-1} (\frac{\partial}{\partial\theta_\alpha}\mathbf{\Sigma}) \mathbf{\Sigma}^{-1}(\frac{\partial}{\partial\theta_\beta}\mathbf{\Sigma})\big],
\end{split}
\end{equation}
with $\alpha,\beta\in\{\tau,\omega\}$.
In practice, however, the receiver does not access the full quadrature vector but only the specific quadrature combinations selected by the detection scheme. In our implementation, the two returned modes are measured by HD. Each mode is interfered with a strong local oscillator (LO), whose phases $(\theta_1,\theta_2)$ respectively, define the quadrature selected by the measurement as depicted in Fig.\ref{fig0}. The detected observables are therefore: 
\begin{equation}
\mathbf{y}=\begin{pmatrix} x_1\cos\theta_1+p_1\sin\theta_1\\ x_2\cos\theta_2+p_2\sin\theta_2
\end{pmatrix}.
\end{equation}
The receiver accesses only a two-dimensional slice of the full four-dimensional Gaussian state covariance matrix. Any information encoded in phase-space directions not selected by the chosen LO phases, does not directly contribute to the observed statistics. 
Parameter estimation must therefore be based on the experimentally accessible outcomes $\mathbf{y}$, so the relevant likelihood is the Gaussian distribution $p(\mathbf{y}|\tau,\omega)$, centered in $\mathbf{\overline{x}}_\text{meas}(\tau,\omega)$ and with covariance $\mathbf{\Sigma}_\text{meas}(\tau,\omega)$.
The optimization of the measurement settings, therefore, plays a central role in the protocol (see Appendix for more details). 
We numerically characterize the joint estimation performance by optimizing the two homodyne phases by computing the product of $F^{-1}_{\tau\tau}F^{-1}_{\omega\omega}$, which quantifies the simultaneous reconstruction precision of the delay-like and Doppler-like parameters.
In the ideal scenario, the local parameter encoding favors nonclassical resources distributed across the two modes, allowing the separable squeezed probe to approach the ultimate precision bound (see Fig.\ref{fig1}\textbf{a}). By contrast, the twin-beam probe redistributes part of the same resource into intermodal correlations, preserving enhanced performance compared to classical coherent probes, but degrading its performance compared to separable squeezed states.

\emph{Noise model --} In practice, the optical field is degraded by propagation loss, background photons, receiver imperfections, and, in adversarial settings, by deliberately injected interference. This is especially relevant in modern LiDAR applications, where spoofing and jamming attacks are now recognized as a genuine systems-level concern, from autonomous sensing to secure ranging. Recent work on quantum-secured LiDAR has reinforced the point that the relevant question is not only whether a quantum probe is advantageous in principle, but whether that advantage survives under physically plausible interference conditions \cite{mrozowski2024demonstration, blakey2022quantum}.

In our model, we separate the disturbance into two conceptually distinct parts. The first is the standard Gaussian sensing channel, accounting for attenuation and thermal background. We consider a channel with transmissivity $\kappa$ and mean number of thermal photons $n_\text{th}$, acting on the first and second order moments as follows:
\begin{equation}
\mathbf{\overline{x}}\xrightarrow{}\sqrt{\kappa}\mathbf{\overline{x}} \qquad \mathbf{\Sigma}\xrightarrow{}\kappa\mathbf{\Sigma}+(1-\kappa)(2 n_\text{th}+1)\mathcal{I}.
\end{equation}
To describe more realistic operating conditions, we then include classical jamming in this setting by modeling the effect of correlated classical disturbance acting on the two returned modes. 
Such correlations can arise from common-mode environmental fluctuations, shared propagation paths, receiver cross-talk, or deliberately injected optical interference with an intermodal structure.
At the receiver, this manifests itself as excess quadrature fluctuations that are fully correlated across the two modes rather than independent.
We model this effect by introducing an additional Gaussian disturbance acting on the detected quadratures at the receiver, described by a covariance  $\Sigma_{jm}(\sigma_c)$ whose strength is set by $\sigma_c$.
Physically, this term represents a structured classical disturbance that produces common excess fluctuations across the two measured modes, rather than independent local noise. It may arise from common-mode environmental fluctuations, shared propagation paths, or deliberately injected optical jamming that couples to both sensing channels:

\begin{equation}
\mathbf{\Sigma}_\text{jm} = \sigma_c^2
    \begin{pmatrix}
        1 & 1 &0& 0\\
        1& 1 &0 &0\\
        0&0 &1&1\\
        0&0&1&1
    \end{pmatrix}.
\end{equation}
This decomposition distinguishes unavoidable propagation noise from structured external interference, allowing us to isolate the effect of correlated intermodal disturbances on the estimation performance.

In the noisy regime, squeezed probes retain a clear advantage over the classical benchmark across a broad range of parameters. In particular, when correlated interference is introduced, having access to an intermodal structure becomes especially relevant. To model this adversarial scenario, we introduce an additive classical covariance matrix $\mathbf{\Sigma}_\text{jm}$ which acts directly on the two-mode Gaussian statistics accessible at the receiver.

It is in this regime that the twin-beam probe becomes a useful resource with respect to both the classical and separable squeezed states as shown in Fig.\ref{fig1}\textbf{a}. For the homodyne scheme considered here, the coherent state variance product increases with the jamming strength $\sigma_c$, whereas both squeezed probes approach a finite plateau. This behavior is a direct consequence of the receiver geometry. For coherent probes, the measured covariance is rotation-invariant and can not restore rank lost under common-mode jamming. By contrast, for squeezed probes, the encoding rotates the squeezed quadrature, leaving a finite covariance response that regularizes the receiver performance even when the mean contribution is strongly suppressed if the measurement angles $(\theta_1, \theta_2)$ are optimized. To have an overall picture of the probes robustness against noise, we report the comparison of performance for different noise levels in Fig.\ref{fig1}\textbf{b-c} by fixing the number of thermal photons to the $1\%$ of the overall signal budget. 

The optimal LO phases depend on both the operating point $(\tau,\omega)$ and the noise level, including loss, thermal background, and correlated jamming. As these quantities vary, the quadrature basis that maximizes the accessible information changes. This makes adaptive receiver strategies especially appealing. In practice, the LO phases can be updated in real time through an online feedback loop, using either prior information or intermediate estimates to track fluctuations, compensate slow drifts, and maintain the receiver close to its optimal working point as developed in \cite{cimini2019adaptive, valeri2023experimental, belliardo2024optimizing, cimini2023deep, minati2025multiparameter}.
Results obtained without receiver adaptation are reported in the Appendix.

\emph{Estimator performance --} A central point of our work is to study whether the presence of squeezing and quantum correlations in the probe translates into an operational advantage at the level of experimentally accessible measurements.

\begin{figure}[htb!]
	\includegraphics[width=0.99\columnwidth]{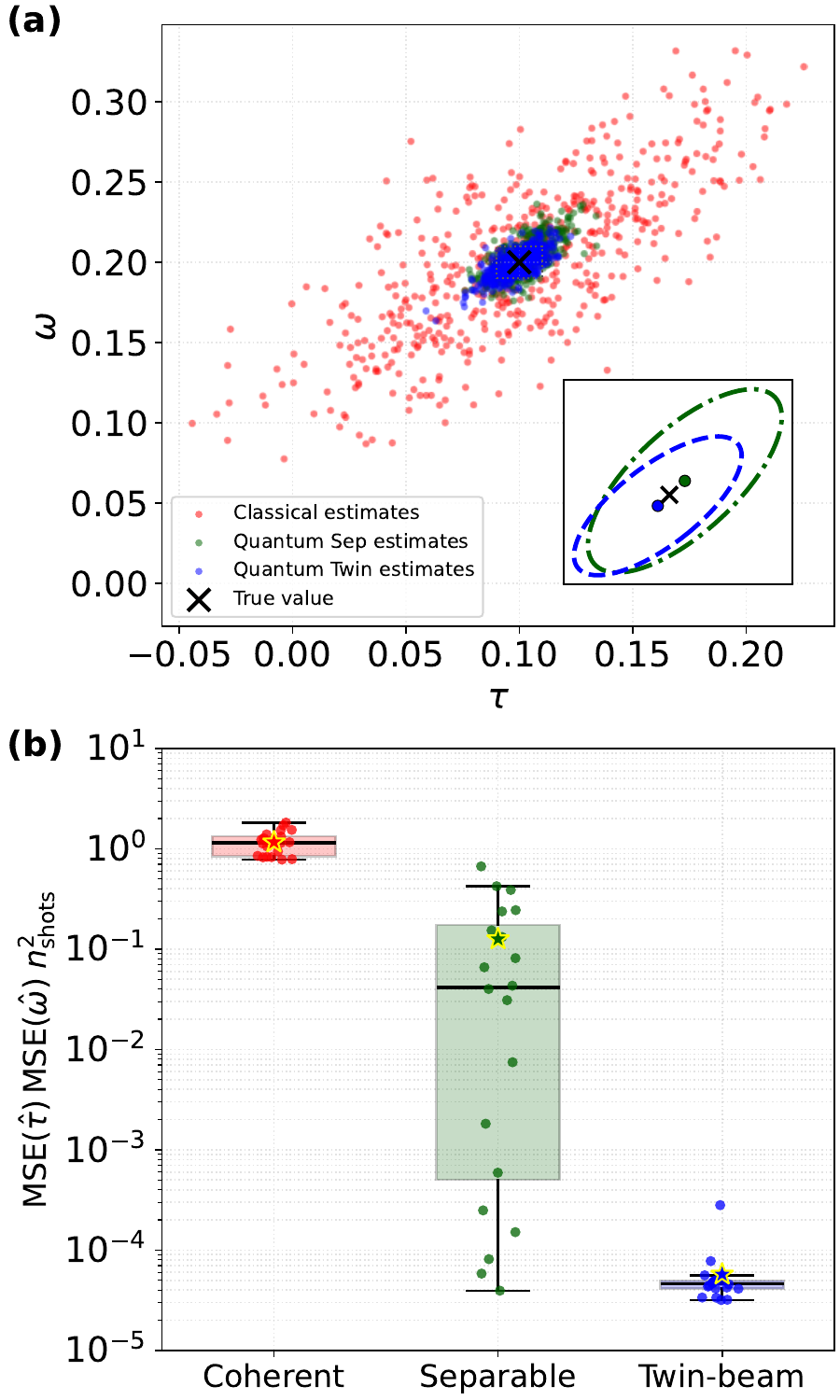}
	\caption{Estimator-level reconstruction from homodyne data for the three probe classes at fixed total photon number $N=50$, $\kappa = 0.95$, $n_\text{th} = 0$, and $\sigma_c=5$: coherent (red), separable squeezed (green), and twin-beam (blue). \textbf{a} Each point in the main panel corresponds to one independent maximum-likelihood estimate $(\hat{\tau},\hat{\omega})$ obtained from a simulated homodyne dataset of fixed size $n_\text{shots} = 100$, generated for the same true operating point $(\tau_\text{true},\omega_\text{true})$ (black cross). The inset shows the corresponding covariance ellipses, centered at the sample mean of each estimator distribution. Their area and orientation quantify the joint uncertainty and the residual correlation between the inferred range- and velocity-like parameters. The twin-beam probe yields the most concentrated estimator cloud and the smallest covariance ellipse. The procedure has been repeated $n_\text{reps} = 500$ times. \textbf{b} Robustness of the estimator-level performance in terms of MSE over 20 random operating points. The distribution of the joint uncertainty obtained from explicit maximum-likelihood reconstruction of the target parameters for the three probe families is reported. The boxes denote the interquartile range, horizontal black lines mark the median, colored stars indicate the mean, and small points represent the individual random operating points.}
	\label{fig2}
\end{figure}

To assess the performance of an explicit estimation strategy, we reconstruct the unknown parameters directly from simulated homodyne data. For a fixed probe, receiver, and operating point $(\tau_\text{true},\omega_\text{true})$, the measured outcomes are described by the Gaussian distribution associated with the experimentally accessible quadratures.
For each configuration, we generate $n_\text{shots}$ independent measurement outcomes $\{\mathbf{y}\}$ so that the simulated data reproduce the homodyne records available in an actual experiment.
The unknown parameters are then estimated via a maximum-likelihood procedure, yielding one parameter pair $(\hat{\tau},\hat{\omega})$.  To quantify the estimator fluctuations, we repeat the full procedure $n_\text{reps}$ times, consisting of generating a new set of homodyne samples and performing an independent maximum-likelihood reconstruction.
From the resulting ensemble of estimates, we compute the sample mean and covariance matrix of the estimator. The diagonal entries of this covariance matrix provide the estimator variances for the delay- and Doppler-like parameters, while the displacement of the sample mean from the true operating point provides a direct check for bias. For a fixed resource budget, the resulting estimator clouds in the $(\tau,\omega)$ plane for a specific parameter setting are reported in Fig.\ref{fig2}\textbf{a}.
This provides a direct estimator-level validation of the receiver-aware metrological model.

For each probe state and noise configuration, we optimize the measurement settings and determine the pair of LO phases $(\theta_{1,opt},\theta_{2,opt})$ that minimizes the variance product $Var(\hat{\tau})Var(\hat{\omega})$ of the inferred parameters. 
The estimator-level analysis is then performed using homodyne data generated with these optimized settings. This ensures that the comparison between probe families is always carried out in the best experimentally accessible homodyne basis for the corresponding operating condition.

We then study the robustness of the estimation protocol versus different parameters values $(\tau_\text{true},\omega_\text{true})$, thus running a robustness study over multiple target configurations.
This study highlights that the observed quantum advantage survives independently of the specific operating point (more details in the Appendix).
We randomly sample $20$ different parameter values. For each sampled point and for each probe family, we optimize the homodyne receiver settings. We then generate simulated homodyne datasets at the chosen operating point, reconstruct the unknown parameters through maximum-likelihood estimation, and repeat the full procedure over $n_\text{reps}=100$ independent Monte Carlo realizations to obtain the empirical estimator statistics.
From the resulting ensemble of reconstructed pairs $(\hat{\tau},\hat{\omega})$, we compute the product of mean-square errors by computing $\langle(\hat{\tau}-\tau_\text{true})^2\rangle\cdot \langle(\hat{\omega}-\omega_\text{true})^2\rangle$ to determine whether performance is degraded by systematic shifts as the operating point is varied. The results of this analysis are reported in Fig.\ref{fig2}\textbf{b}.
The coherent probe systematically yields the largest joint uncertainty, while both quantum probes retain a clear advantage. More specifically, the twin-beam probe displays the smallest distribution of estimator-level uncertainty over the considered operating points, showing that its advantage is not restricted to a finely tuned parameter value, but it is maintained under moderate variations of the target parameters. 
This demonstrates that the advantage enabled by EPR correlations remains operationally accessible in realistic scenarios and at the level of actual homodyne-based parameter reconstruction.

\emph{Conclusion --} We have established a receiver-aware, jamming-resilient framework for
quantum-enhanced LiDAR in the joint estimation of target range and Doppler shift. We moved beyond ideal covariance-level analyses by carrying out explicit simulations of homodyne measurement records employing measurement optimization followed by maximum-likelihood reconstruction of the unknown parameters. In this way, the reported advantage is not inferred only from the information contained in the probe state in principle, but remains accessible at the level of experimentally implementable measurements and estimators.

Within this framework, we show that local squeezing already provides a robust advantage over coherent light and preserves enhanced performance in realistic noisy conditions. Importantly, in the presence of correlated jamming noise, twin-beam probes outperform both classical and separable squeezed strategies, provided that the measurement basis is properly adapted to the operating point and to the noise structure. 
This identifies a concrete regime in which EPR correlations become the key metrological resource and in which entanglement delivers a practical advantage beyond what can be achieved by local squeezing alone for remote sensing applications.

\emph{Acknowledgments --}

We thank Marco Barbieri and Fabio Sciarrino for useful discussion. 
This work is supported by PNRR MUR project PE0000023-NQSTI (Spoke 4).

\vspace{20em}

\bibliography{biblio}

@PREAMBLE{
 "\providecommand{\noopsort}[1]{}" 
 # "\providecommand{\singleletter}[1]{#1}%" 
}

@article{mrozowski2024demonstration,
  title={Demonstration of quantum-enhanced rangefinding robust against classical jamming},
  author={Mrozowski, Mateusz P and Murchie, Richard J and Jeffers, John and Pritchard, Jonathan D},
  journal={Optics Express},
  volume={32},
  number={3},
  pages={2916--2928},
  year={2024},
  publisher={Optica Publishing Group}
}

@article{reichert2024heisenberg,
  title={Heisenberg-limited quantum lidar for joint range and velocity estimation},
  author={Reichert, Maximilian and Zhuang, Quntao and Sanz, Mikel},
  journal={Physical Review Letters},
  volume={133},
  number={13},
  pages={130801},
  year={2024},
  publisher={APS}
}

@article{caves1981quantum,
  title={Quantum-mechanical noise in an interferometer},
  author={Caves, Carlton M},
  journal={Physical Review D},
  volume={23},
  number={8},
  pages={1693},
  year={1981},
  publisher={APS}
}

@article{mcmanamon2012review,
  title={Review of ladar: a historic, yet emerging, sensor technology with rich phenomenology},
  author={McManamon, Paul F},
  journal={Optical Engineering},
  volume={51},
  number={6},
  pages={060901--060901},
  year={2012},
  publisher={Society of Photo-Optical Instrumentation Engineers}
}

@article{wang2021challenges,
  title={Challenges and opportunities in Lidar remote sensing},
  author={Wang, Zhien and Menenti, Massimo},
  journal={Frontiers in Remote Sensing},
  volume={2},
  pages={641723},
  year={2021},
  publisher={Frontiers Media SA}
}

@article{zhang2015entanglement,
  title={Entanglement-enhanced sensing in a lossy and noisy environment},
  author={Zhang, Zheshen and Mouradian, Sara and Wong, Franco NC and Shapiro, Jeffrey H},
  journal={Physical review letters},
  volume={114},
  number={11},
  pages={110506},
  year={2015},
  publisher={APS}
}

@article{li2025noise,
  title={Noise-tolerant LiDAR approaching the standard quantum-limited precision},
  author={Li, Haochen and Zheng, Kaimin and Ge, Rui and Zhang, Labao and Zhang, Lijian and He, Weiji and Zhang, Biao and Wu, Miao and Wang, Ben and Mi, Minghao and others},
  journal={Light: Science \& Applications},
  volume={14},
  number={1},
  pages={138},
  year={2025},
  publisher={Nature Publishing Group UK London}
}

@article{zhuang2017entanglement,
  title={Entanglement-enhanced lidars for simultaneous range and velocity measurements},
  author={Zhuang, Quntao and Zhang, Zheshen and Shapiro, Jeffrey H},
  journal={Physical Review A},
  volume={96},
  number={4},
  pages={040304},
  year={2017},
  publisher={APS}
}

@article{xu2021experimental,
  title={Experimental quantum target detection approaching the fundamental Helstrom limit},
  author={Xu, Feixiang and Zhang, Xiao-Ming and Xu, Liang and Jiang, Tao and Yung, Man-Hong and Zhang, Lijian},
  journal={Physical Review Letters},
  volume={127},
  number={4},
  pages={040504},
  year={2021},
  publisher={APS}
}

@article{comeron2017current,
  title={Current research in lidar technology used for the remote sensing of atmospheric aerosols},
  author={Comer{\'o}n, Adolfo and Mu{\~n}oz-Porcar, Constantino and Rocadenbosch, Francesc and Rodr{\'\i}guez-G{\'o}mez, Alejandro and Sicard, Micha{\"e}l},
  journal={Sensors},
  volume={17},
  number={6},
  pages={1450},
  year={2017},
  publisher={MDPI}
}

@article{northend1966laser,
  title={Laser radar (lidar) for meteorological observations},
  author={Northend, CA and Honey, RC and Evans, WE},
  journal={Review of Scientific Instruments},
  volume={37},
  number={4},
  pages={393--400},
  year={1966}
}

@book{dong2017lidar,
  title={LiDAR remote sensing and applications},
  author={Dong, Pinliang and Chen, Qi},
  year={2017},
  publisher={Crc Press}
}

@article{blakey2022quantum,
  title={Quantum and non-local effects offer over 40 dB noise resilience advantage towards quantum lidar},
  author={Blakey, Phillip S and Liu, Han and Papangelakis, Georgios and Zhang, Yutian and L{\'e}ger, Zacharie M and Iu, Meng Lon and Helmy, Amr S},
  journal={Nature communications},
  volume={13},
  number={1},
  pages={5633},
  year={2022},
  publisher={Nature Publishing Group UK London}
}

@article{chen2024optimal,
  title={Optimal probe States for single-mode quantum target detection in arbitrary object reflectivity},
  author={Chen, Wei-Ming and Tsai, Pin-Ju},
  journal={Physical Review Research},
  volume={6},
  number={2},
  pages={023084},
  year={2024},
  publisher={APS}
}

@article{spedalieri2021optimal,
  title={Optimal squeezing for quantum target detection},
  author={Spedalieri, Gaetana and Pirandola, Stefano},
  journal={Physical Review Research},
  volume={3},
  number={4},
  pages={L042039},
  year={2021},
  publisher={APS}
}

@article{amann2001laser,
  title={Laser ranging: a critical review of unusual techniques for distance measurement},
  author={Amann, Markus-Christian and Bosch, Thierry M and Lescure, Marc and Myllylae, Risto A and Rioux, Marc},
  journal={Optical engineering},
  volume={40},
  pages={10--19},
  year={2001}
}

@article{royo2019overview,
  title={An overview of lidar imaging systems for autonomous vehicles},
  author={Royo, Santiago and Ballesta-Garcia, Maria},
  journal={Applied sciences},
  volume={9},
  number={19},
  pages={4093},
  year={2019},
  publisher={MDPI}
}

@article{nichols2018multiparameter,
  title={Multiparameter Gaussian quantum metrology},
  author={Nichols, Rosanna and Liuzzo-Scorpo, Pietro and Knott, Paul A and Adesso, Gerardo},
  journal={Physical Review A},
  volume={98},
  number={1},
  pages={012114},
  year={2018},
  publisher={APS}
}

@article{bressanini2024multi,
  title={Multi-parameter quantum estimation of single-and two-mode pure Gaussian states},
  author={Bressanini, Gabriele and Genoni, Marco G and Kim, MS and Paris, Matteo GA},
  journal={Journal of Physics A: Mathematical and Theoretical},
  volume={57},
  number={31},
  pages={315305},
  year={2024},
  publisher={IOP Publishing}
}

@article{jo2021quantum,
  title={Quantum illumination receiver using double homodyne detection},
  author={Jo, Yonggi and Lee, Sangkyung and Ihn, Yong Sup and Kim, Zaeill and Lee, Su-Yong},
  journal={Physical Review Research},
  volume={3},
  number={1},
  pages={013006},
  year={2021},
  publisher={APS}
}

@article{reichert2023quantum,
  title={Quantum illumination with a hetero-homodyne receiver and sequential detection},
  author={Reichert, Maximilian and Zhuang, Quntao and Shapiro, Jeffrey H and Di Candia, Roberto},
  journal={Physical Review Applied},
  volume={20},
  number={1},
  pages={014030},
  year={2023},
  publisher={APS}
}

@article{zhuang2021quantum,
  title={Quantum ranging with Gaussian entanglement},
  author={Zhuang, Quntao},
  journal={Physical Review Letters},
  volume={126},
  number={24},
  pages={240501},
  year={2021},
  publisher={APS}
}

@article{tan2008quantum,
  title={Quantum illumination with Gaussian states},
  author={Tan, Si-Hui and Erkmen, Baris I and Giovannetti, Vittorio and Guha, Saikat and Lloyd, Seth and Maccone, Lorenzo and Pirandola, Stefano and Shapiro, Jeffrey H},
  journal={Physical review letters},
  volume={101},
  number={25},
  pages={253601},
  year={2008},
  publisher={APS}
}

@article{lloyd2008enhanced,
  title={Enhanced sensitivity of photodetection via quantum illumination},
  author={Lloyd, Seth},
  journal={Science},
  volume={321},
  number={5895},
  pages={1463--1465},
  year={2008},
  publisher={American Association for the Advancement of Science}
}

@article{torrome2024advances,
  title={Advances in quantum radar and quantum LiDAR},
  author={Torrom{\'e}, Ricardo Gallego and Barzanjeh, Shabir},
  journal={Progress in Quantum Electronics},
  volume={93},
  pages={100497},
  year={2024},
  publisher={Elsevier}
}

@article{shapiro2009quantum,
  title={Quantum illumination versus coherent-state target detection},
  author={Shapiro, Jeffrey H and Lloyd, Seth},
  journal={New Journal of Physics},
  volume={11},
  number={6},
  pages={063045},
  year={2009}
}

@article{sorelli2021detecting,
  title={Detecting a target with quantum entanglement},
  author={Sorelli, Giacomo and Treps, Nicolas and Grosshans, Fr{\'e}d{\'e}ric and Boust, Fabrice},
  journal={IEEE Aerospace and Electronic Systems Magazine},
  volume={37},
  number={5},
  pages={68--90},
  year={2021},
  publisher={IEEE}
}

@article{huang2021quantum,
  title={Quantum-limited estimation of range and velocity},
  author={Huang, Zixin and Lupo, Cosmo and Kok, Pieter},
  journal={PRX Quantum},
  volume={2},
  number={3},
  pages={030303},
  year={2021},
  publisher={APS}
}

@article{pirandola2018advances,
  title={Advances in photonic quantum sensing},
  author={Pirandola, Stefano and Bardhan, B Roy and Gehring, Tobias and Weedbrook, Christian and Lloyd, Seth},
  journal={Nature Photonics},
  volume={12},
  number={12},
  pages={724--733},
  year={2018},
  publisher={Nature Publishing Group UK London}
}

@article{karsa2024quantum,
  title={Quantum illumination and quantum radar: A brief overview},
  author={Karsa, Athena and Fletcher, Alasdair and Spedalieri, Gaetana and Pirandola, Stefano},
  journal={Reports on progress in physics},
  volume={87},
  number={9},
  pages={094001},
  year={2024},
  publisher={IOP Publishing}
}

@article{cimini2024benchmarking,
  title={Benchmarking Bayesian quantum estimation},
  author={Cimini, Valeria and Polino, Emanuele and Valeri, Mauro and Spagnolo, Nicolo and Sciarrino, Fabio},
  journal={Quantum Science and Technology},
  volume={9},
  number={3},
  pages={035035},
  year={2024},
  publisher={IOP Publishing}
}

@article{belliardo2024optimizing,
  title={Optimizing quantum-enhanced Bayesian multiparameter estimation of phase and noise in practical sensors},
  author={Belliardo, Federico and Cimini, Valeria and Polino, Emanuele and Hoch, Francesco and Piccirillo, Bruno and Spagnolo, Nicol{\`o} and Giovannetti, Vittorio and Sciarrino, Fabio},
  journal={Physical Review Research},
  volume={6},
  number={2},
  pages={023201},
  year={2024},
  publisher={APS}
}

@article{cimini2024variational,
  title={Variational quantum algorithm for experimental photonic multiparameter estimation},
  author={Cimini, Valeria and Valeri, Mauro and Piacentini, Simone and Ceccarelli, Francesco and Corrielli, Giacomo and Osellame, Roberto and Spagnolo, Nicol{\`o} and Sciarrino, Fabio},
  journal={npj Quantum Information},
  volume={10},
  number={1},
  pages={26},
  year={2024},
  publisher={Nature Publishing Group UK London}
}

@article{cimini2019adaptive,
  title={Adaptive tracking of enzymatic reactions with quantum light},
  author={Cimini, Valeria and Mellini, Marta and Rampioni, Giordano and Sbroscia, Marco and Leoni, Livia and Barbieri, Marco and Gianani, Ilaria},
  journal={Optics Express},
  volume={27},
  number={24},
  pages={35245--35256},
  year={2019},
  publisher={Optical Society of America}
}

@article{valeri2023experimental,
  title={Experimental multiparameter quantum metrology in adaptive regime},
  author={Valeri, Mauro and Cimini, Valeria and Piacentini, Simone and Ceccarelli, Francesco and Polino, Emanuele and Hoch, Francesco and Bizzarri, Gabriele and Corrielli, Giacomo and Spagnolo, Nicol{\`o} and Osellame, Roberto and others},
  journal={Physical Review Research},
  volume={5},
  number={1},
  pages={013138},
  year={2023},
  publisher={APS}
}

@article{cimini2023deep,
  title={Deep reinforcement learning for quantum multiparameter estimation},
  author={Cimini, Valeria and Valeri, Mauro and Polino, Emanuele and Piacentini, Simone and Ceccarelli, Francesco and Corrielli, Giacomo and Spagnolo, Nicol{\`o} and Osellame, Roberto and Sciarrino, Fabio},
  journal={Advanced Photonics},
  volume={5},
  number={1},
  pages={016005},
  year={2023},
  publisher={SPIE}
}

@article{roccia2018multiparameter,
  title={Multiparameter approach to quantum phase estimation with limited visibility},
  author={Roccia, Emanuele and Cimini, Valeria and Sbroscia, Marco and Gianani, Ilaria and Ruggiero, Ludovica and Mancino, Luca and Genoni, Marco G and Ricci, Maria Antonietta and Barbieri, Marco},
  journal={Optica},
  volume={5},
  number={10},
  pages={1171--1176},
  year={2018},
  publisher={Optical Society of America}
}

@article{minati2025multiparameter,
  title={Multiparameter quantum-enhanced adaptive metrology with squeezed light},
  author={Minati, Giorgio and Urbani, Enrico and Spagnolo, Nicol{\`o} and Cimini, Valeria and Sciarrino, Fabio},
  journal={arXiv preprint arXiv:2510.14739},
  year={2025}
}

@article{moreva2025quantum,
  title={Quantum photonics sensing in biosystems},
  author={Moreva, Ekaterina and Cimini, Valeria and Gianani, Ilaria and Bernardi, Ettore and Traina, Paolo and Degiovanni, Ivo P and Barbieri, Marco},
  journal={APL Photonics},
  volume={10},
  number={1},
  year={2025},
  publisher={AIP Publishing}
}

@article{manrique2026quantum,
  title={Quantum noise in ranging with optical pulses},
  author={Manrique, Mylenne and Gianani, Ilaria and Barbieri, Marco and Parigi, Valentina and Treps, Nicolas},
  journal={arXiv preprint arXiv:2604.05107},
  year={2026}
}

@article{lawrie2019quantum,
  title={Quantum sensing with squeezed light},
  author={Lawrie, Benjamin J and Lett, Paul D and Marino, Alberto M and Pooser, Raphael C},
  journal={Acs Photonics},
  volume={6},
  number={6},
  pages={1307--1318},
  year={2019},
  publisher={ACS Publications}
}

@article{giovannetti2001quantum,
  title={Quantum-enhanced positioning and clock synchronization},
  author={Giovannetti, Vittorio and Lloyd, Seth and Maccone, Lorenzo},
  journal={Nature},
  volume={412},
  number={6845},
  pages={417--419},
  year={2001},
  publisher={Nature Publishing Group UK London}
}

@article{polino2020photonic,
  title={Photonic quantum metrology},
  author={Polino, Emanuele and Valeri, Mauro and Spagnolo, Nicol{\`o} and Sciarrino, Fabio},
  journal={AVS Quantum Science},
  volume={2},
  number={2},
  year={2020},
  publisher={AIP Publishing}
}

@article{barbieri2022optical,
  title={Optical quantum metrology},
  author={Barbieri, Marco},
  journal={Prx Quantum},
  volume={3},
  number={1},
  pages={010202},
  year={2022},
  publisher={APS}
}

@article{taylor2016quantum,
  title={Quantum metrology and its application in biology},
  author={Taylor, Michael A and Bowen, Warwick P},
  journal={Physics Reports},
  volume={615},
  pages={1--59},
  year={2016},
  publisher={Elsevier}
}

@article{petit2015remote,
  title={Remote attacks on automated vehicles sensors: Experiments on camera and lidar},
  author={Petit, Jonathan and Stottelaar, Bas and Feiri, Michael and Kargl, Frank},
  journal={Black Hat Europe},
  volume={11},
  number={2015},
  pages={995},
  year={2015},
  publisher={Amsterdam, Netherlands}
}

@inproceedings{cao2019adversarial,
  title={Adversarial sensor attack on lidar-based perception in autonomous driving},
  author={Cao, Yulong and Xiao, Chaowei and Cyr, Benjamin and Zhou, Yimeng and Park, Won and Rampazzi, Sara and Chen, Qi Alfred and Fu, Kevin and Mao, Z Morley},
  booktitle={Proceedings of the 2019 ACM SIGSAC conference on computer and communications security},
  pages={2267--2281},
  year={2019}
}

%\emph{Data availability --}
\section*{Appendices}

\subsection*{General parameter encodings}

\begin{figure}[htb!]
	\includegraphics[width=0.99\columnwidth]{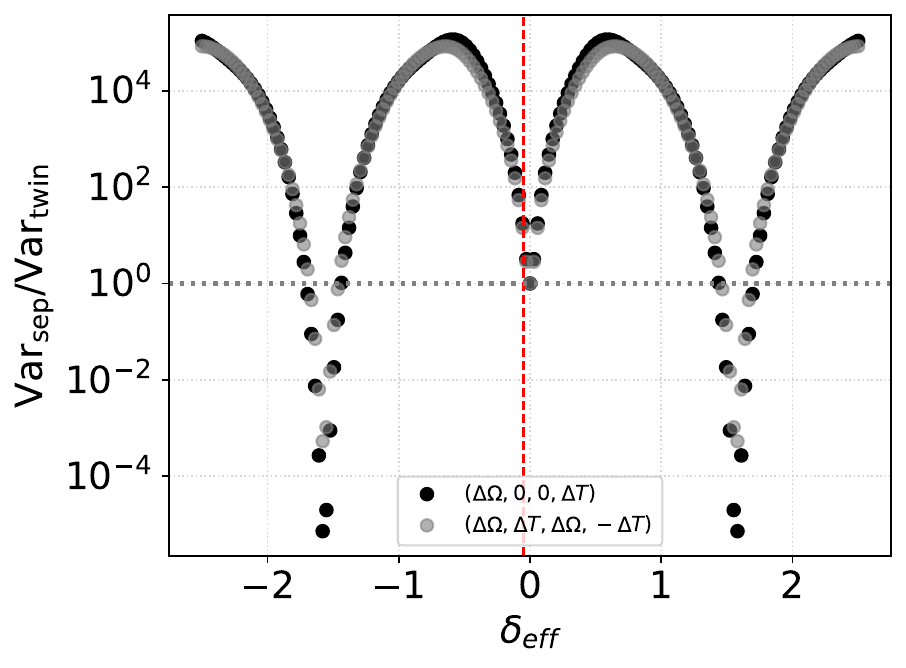}
	\caption{State-level robustness of the twin-beam advantage under generalized two-mode encoding. We plot the variance product ratio, evaluated from the complete two-mode Gaussian covariance matrix and mean vector, for $\sigma_c = 100$, $\kappa=1$, and $n_\text{th}=0$ as a function of the encoding asymmetry $\delta_{\rm eff}$. Black markers correspond to the independent encoding used in the main text, while gray markers correspond to the mixed encoding. The red dashed line marks the operating point used in the first part of the main-text analysis.}
	\label{app0}
\end{figure}

The structural distinction between separable squeezing and twin-beam entanglement under common-mode jamming is robust against the specific choice of linear two-mode encoding. To assess this, we consider the general local phase map:
\begin{equation}
R_{1}\!\bigl(a\tau+b\omega\bigr)\otimes
R_{2}\!\bigl(c\tau+d\omega\bigr),
\qquad
\det\!\begin{pmatrix}a&b\\c&d\end{pmatrix}\neq 0,
\label{eq:bilat_enc}
\end{equation}
which is the most general invertible linear encoding that acts via independent phase-space rotations on the two signal modes. The encoding used in the main text corresponds to the special case $(a,b,c,d)=(\Delta\Omega,0,0,\Delta T)$.
Defining:
\begin{equation}
\delta_1=a\tau+b\omega,\qquad
\delta_2=c\tau+d\omega,\qquad
\delta_{\rm eff}=\frac{\delta_1-\delta_2}{2},
\end{equation}
for this comparison, we evaluate the achieved performance considering the full Gaussian state rather than a specific receiver projection. Concretely, for each encoding, we compute the complete two-mode covariance matrix and mean vector after the parameter-dependent rotations, and from them evaluate the full Fisher information matrix (FIM). This allows us to isolate the effect of the encoding, independently of receiver optimization. More specifically, we compare two distinct bilateral encodings, the one used in the main text, and the encoding $(a,b,c,d)=(\Delta\Omega,\Delta T,\Delta\Omega,-\Delta T)$, in which both modes acquire the same delay contribution but opposite Doppler signs. As shown in Fig.~\ref{app0}, the numerical results show the same trend when plotted as a function of $\delta_{\rm eff}$. This shows that the state-level twin-beam advantage is controlled by the effective relative encoding angle $\delta_{\rm eff}$, rather than by the specific assignment of the parameters to the two modes.
The condition $ad-cb\neq0$ ensures joint identifiability of $(\tau,\omega)$, while the degenerate case $a=c$ and $b=d$ is
excluded, since it would make the two modes carry the same encoded phase and therefore destroy independent parameter reconstruction.

\subsection*{Adaptive homodyne measurement angles}

A central aspect of the present protocol is that the receiver is not fixed a priori, but it can adapt its measurement basis depending on the actual operating conditions. This is essential because the amount of information that is operationally accessible depends on the quadratures selected by the measurement.
The optimization variables are the LO phases $\theta_1$ and $\theta_2$, which determine the quadratures measured on the two reflected modes.

\begin{figure}[htb!]
	\includegraphics[width=0.99\columnwidth]{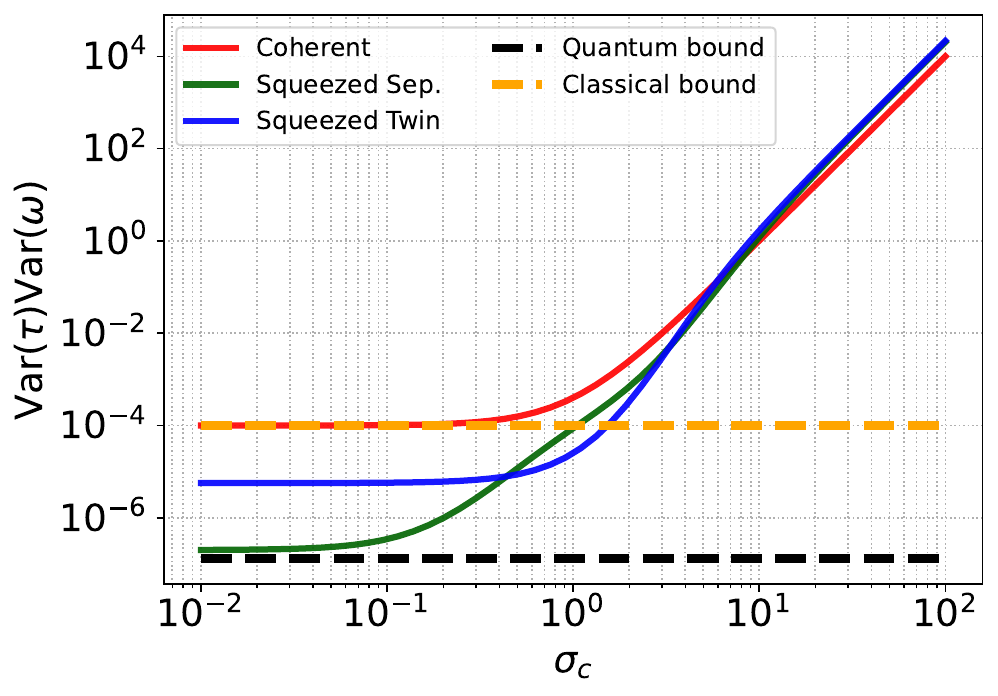}
	\caption{Product of the estimator variances as a function of the correlated-noise strength $\sigma_c$ for coherent (red), separable squeezed (green), and twin-beam (blue) probes for a fixed homodyne basis.}
	\label{app1}
\end{figure}

The receiver optimization is carried out locally at the operating point $(\tau_\text{true},\omega_\text{true})$ by minimizing the product of the diagonal elements of the inverse of the FIM. This quantifies the joint uncertainty associated with the reconstruction of the delay-like and Doppler-like parameters and provides a natural objective function for the adaptive tuning of the receiver. For each choice of receiver parameters, we construct the corresponding Gaussian measurement model and evaluate the FIM at the operating point. Because the optimization landscape can be highly nonconvex, we perform the search numerically using the Tree-structured Parzen Estimator (TPE) algorithm implemented in Python package \emph{Optuna}, with $n_\text{trials}$ explored receiver configurations. Importantly, the FIM entering this cost function is computed from the Gaussian likelihood of the actually measured outcomes, $p(\mathbf{y}|\tau,\omega)$. The optimization thus identifies the measurement basis that maximizes the information that is operationally accessible, rather than the information contained in the full state in principle.

To highlight the importance of receiver optimization, Fig.~\ref{app1} compares the performances reported in Fig.~2\textbf{a} of the main text with those obtained from a pre-calibrated receiver, where the two LO phases are frozen at their optimal values in the ideal scenario. This comparison shows that for high values of correlated noise, if the measurement is not properly optimized, the advantage of both the separable and twin-beam probes is lost.

\end{document}